\pdfoutput=1
\documentclass{article}
\usepackage{natbib}
\usepackage[colorlinks=true, citecolor=blue, urlcolor=black]{hyperref}
\usepackage{graphicx}
\usepackage{amssymb}
\usepackage{amsmath}
\usepackage{color,soul,cancel,subfigure}
\usepackage{setspace,mathrsfs,soul,cancel}
\usepackage{xargs}

\def\bzero{{\bf 0}}

\def\ba{{\mbox{\boldmath$a$}}}

\def\bm{{\bf m}}

\def\bq{{\bf q}}
\def\br{{\bf r}}

\def\bt{{\bf t}}
\def\bu{{\bf u}}
\def\bv{{\bf v}}
\def\bw{{\bf w}}
\def\bx{{\bf x}}
\def\by{{\bf y}}

\def\bV{{\bf V}}

\def\thick#1{\hbox{\rlap{$#1$}\kern0.25pt\rlap{$#1$}\kern0.25pt$#1$}}
\def\balpha{\boldsymbol{\alpha}}
\def\bbeta{\boldsymbol{\beta}}

\def\bvarepsilon{\boldsymbol{\varepsilon}}

\def\btheta{\boldsymbol{\theta}}

\def\bmu{\boldsymbol{\mu}}
\def\bnu{\boldsymbol{\nu}}
\def\bxi{\boldsymbol{\xi}}

\def\bphi{\boldsymbol{\phi}}

\def\smbalpha{\boldsymbol{{\scriptstyle{\alpha}}}}

\def\ahat{{\widehat a}}

\def\mhat{{\widehat m}}

\def\phat{{\widehat p}}

\def\vhat{{\widehat v}}

\def\Qhat{{\widehat Q}}

\def\atilde{{\widetilde a}}
\def\btilde{{\widetilde b}}

\def\Stilde{{\widetilde S}}

\def\bmhat{{\widehat \bm}}

\def\bqhat{{\widehat \bq}}

\def\bqtilde{{\widetilde \bq}}

\def\bxtilde{{\widetilde \bx}}

\def\alphahat{{\widehat\alpha}}

\def\omegahat{{\widehat\omega}}

\def\Sigmahat{{\widehat\Sigma}}

\def\epsilontilde{{\widetilde\epsilon}}

\def\omegatilde{{\widetilde\omega}}

\def\balphahat{{\widehat\balpha}}
\def\bbetahat{{\widehat\bbeta}}

\def\bmuhat{{\widehat\bmu}}

\def\bxihat{{\widehat\bxi}}

\def\smbalpha{\widehat{\smbalpha}}

\def\hbar{\bar{ h}}

\def\Isc{{\cal I}}

\def\Lsc{{\cal L}}

\def\Qsc{{\cal Q}}

\def\Tsc{{\cal T}}
\def\Usc{{\cal U}}

\def\Tschat{\widehat{{\cal T}}}

\def\transpose{{\sf \scriptscriptstyle{T}}}

\def\half{\frac{1}{2}}

\def\nnhalf{n^{-\half}}

\def\var{\mbox{var}}

\def\sumin{\sum_{i=1}^n}

\def\trans{^{\transpose}}
\def\inv{^{-1}}

\def\cov{\mbox{cov}}

\def\var{\mbox{var}}

\def\mybox#1{\vskip1mm \begin{center}
        \hspace{.0\textwidth}\vbox{\hrule\hbox{\vrule\kern6pt
\parbox{.9\textwidth}{\kern6pt#1\vskip6pt}\kern6pt\vrule}\hrule}
        \end{center} \vskip-5mm}
\def\lboxit#1{\vbox{\hrule\hbox{\vrule\kern6pt
      \vbox{\kern6pt#1\vskip6pt}\kern6pt\vrule}\hrule}}

\def\thickboxit#1{\vbox{{\hrule height 1mm}\hbox{{\vrule width 1mm}\kern6pt
          \vbox{\kern6pt#1\kern6pt}\kern6pt{\vrule width 1mm}}
               {\hrule height 1mm}}}

\def\fat#1{\hbox{\rlap{$#1$}\kern0.25pt\rlap{$#1$}\kern0.25pt$#1$}}

\def\sumin{\sum_{i=1}^n}

\def\inv{^{-1}} 	
\def\Tsc{\mathcal T}
\def\Isc{\mathcal I}

\def\var{\text{Var}}
\def\cov{\text{Cov}}

\def\Vbb{\mathbb{V}}

\def\Usc{\mathcal{U}}

\begin{document}

\title{Variance component score test for time-course gene set analysis of longitudinal RNA-seq data}

\author{DENIS AGNIEL\\
\small{\textit{Department of Biomedical Informatics, Harvard Medical School, Boston, MA, USA}}\normalsize\medskip\\
BORIS P. HEJBLUM
	\thanks{bhejblum@hsph.harvard.edu}\\
\small{\textit{Department of Biostatistics, Harvard T.H. Chan School of Public Health, Boston, MA, USA}}\\
\small{\textit{University of Bordeaux, ISPED, INSERM U1219, INRIA SISTM, Bordeaux, FRANCE}}\\
\small{\textit{Vaccine Research Institute, Cr\'eteil, FRANCE}}\normalsize
}

\markboth%
{D. Agniel and B. Hejblum}
{Time-course RNA-seq gene set analysis}

\maketitle


\begin{abstract}
{As gene expression measurement technology is shifting from microarrays to sequencing, the statistical tools available for their analysis must be adapted since RNA-seq data are measured as counts. Recently, it has been proposed to tackle the count nature of these data by modeling log-count reads per million as continuous variables, using nonparametric regression to account for their inherent heteroscedasticity. Adopting such a framework, we propose tcgsaseq, a principled, model-free and efficient top-down method for detecting longitudinal changes in RNA-seq gene sets. Considering gene sets defined \textit{a priori}, tcgsaseq identifies those whose expression vary over time, based on an original variance component score test accounting for both covariates and heteroscedasticity without assuming any specific parametric distribution for the transformed counts. We demonstrate that despite the presence of a nonparametric component, our test statistic has a simple form and limiting distribution, and both may be computed quickly. A permutation version of the test is additionally proposed for very small sample sizes. Applied to both simulated data and two real datasets, the proposed  method is shown to exhibit very good statistical properties, with an increase in stability and power when compared to state of the art methods ROAST, edgeR and DESeq2, which can fail to control the type I error under certain realistic settings. We have made the method available for the community in the R package \texttt{tcgsaseq}.
}\bigskip\medskip\bigskip\smallskip
\end{abstract}

\section{Introduction}

Gene expression is a dynamic biological process of living organisms, whose dysfunction and variation can be related to numerous diseases. During the past two decades, gene expression measurements have developed rapidly, thanks to wide dissemination of microarray technology. In recent years, gene expression measurement technology has been shifting from microarrays to sequencing (RNA-seq) technology. The higher resolution of RNA-seq technology provides a number of advantages over microarrays, among which are the ability to make \textit{de novo} discoveries and an increased sensitivity to low abundance variants \citep{Marioni2008a, Malone2011a}. Indeed, RNA-seq is not restricted to a predefined set of probes like microarrays but can instead measure the genome in its entirety.

A large body of statistical methods have been developed to analyze microarray data. But as technology for measuring gene expression is transitioning to RNA-seq, new methodological challenges arise. RNA-seq produces count data, while microarray analysis techniques generally assume continuity. Due to their underlying count nature, RNA-seq data are intrinsically heteroscedastic. Various approaches have been proposed to deal with these issues, mostly relying on modeling the underlying count nature of the data through the use of Poisson or negative binomial distributions \citep{Marioni2008a, Simon2010, Robinson2010}. Recently, \cite{Law2014} have instead proposed to use normal-based methods to analyze RNA-seq data by explicitly modeling the heteroscedasticity and accounting for it by weighting. 

While most methods for gene expression data focus on univariate differential gene expression analysis, it has been shown that gene set analysis (GSA) can be a more powerful and interpretable alternative \citep{Subramanian2005, Hejblum2015a}. GSA uses \textit{a priori} defined gene sets annotated with biological functions and investigates their potential association with biological conditions of interest. Typically, GSA is defined by the type of hypothesis tested and the method of aggregating information across genes. There are two primary approaches to specifying hypotheses: self-contained hypotheses and competitive hypotheses \citep{Goeman2007b}. \cite{Rahmatallah2015} recently showed that self-contained tests tend to be more powerful and more robust than competitive ones. Furthermore, some GSA tests rely on gene-level univariate statistics as a first step, using a bottom-up enrichment approach. However, when signal strength is weak, single-step, top-down approaches relying on multivariate modeling better leverage the additional power of GSA \citep{Hejblum2015a}.

Another concern, particularly in the longitudinal setting, is adequately accounting for heterogeneity of effects \citep{Cui2016}. GSA approaches often overlook the potential longitudinal heterogeneity within a gene set while such heterogeneity is not infrequent \citep{Ackermann2009} and can be of biological interest \citep{Prieto2006, Hu2013}, especially if considered gene sets correspond to biological pathways. \citet{Hejblum2015a} showed the potential statistical power gain by accounting for this heterogeneity in longitudinal microarray studies.

As costs keep decreasing for RNA-seq experiments, more complex study designs, such as time-course experiments, have become more common \citep{Nueda2014, Dorr2015, Leng2015, Baduel2016}. However, very few GSA approaches can properly accommodate and test hypotheses more complex than simple differential expression, such as change over time. The ROAST method \citep{Wu2010a}, which is a linear-model-based testing procedure that can make use of the weighting in \cite{Law2014}, has been identified in the recent review by \cite{Rahmatallah2015} as one of the top-performing GSA methods for RNA-seq data. Among the eleven methods compared in the review, only three are derived from top-down models incorporating all the genes in the set at once, and ROAST is the only one that is not limited to comparisons of two groups. Additionally, DESeq2 \citep{Love2014} and edgeR \citep{Robinson2010, McCarthy2012}, are currently the most prominent approaches used for gene level differential analysis of RNA-seq data. They both rely on the assumption that gene counts from RNA-seq measurements follow a negative binomial distribution. edgeR uses the ROAST method to propose a built-in framework for self-contained gene set testing, while DESeq2 can only perform gene-wise tests.

In this article, we propose {\em tcgsaseq}, a method to analyze RNA-seq data at the gene-set level, with a particular focus on longitudinal studies. We derive a variance component score test, similar to those that have been proposed in other testing situations \citep{
Wu2011b, Huang2013c}, that facilitates testing both homogeneous and heterogeneous gene sets simultaneously. Variance component tests offer the speed and simplicity of standard score tests, but potentially gain statistical power by using many fewer degrees of freedom. Inspired by the voom approach \citep{Law2014}, we propose to estimate the mean-variance relationship in a more principled way using local linear regression, to account for the inherent heteroscedasticity of the data. Despite this nonparametric step, we demonstrate that the test statistic has a simple limiting distribution that is valid regardless of any model specification. We also propose a permutation version of the test to deal with small sample sizes. Our method is implemented in the R package \texttt{tcgsaseq}, available on the Comprehensive R Archive Network at \url{https://cran.r-project.org/web/packages/tcgsaseq}.

Our general approach to GSA in longitudinal RNA-seq studies has three primary advantages over existing approaches. First, unlike ROAST, our variance component approach remains valid under model misspecification and does not rely on {\em ad hoc} aggregation of information across genes. Second, unlike previous approaches to variance component testing in microarray data \citep{Huang2013c}, our approach can accommodate the intrinsic mean-variance relationship in RNA-seq data, while remaining fast to compute. Third, our test remains powerful even when patients display heterogeneous trajectories over time. 

The remainder of the paper is organized as follows. Section \ref{method} describes our variance component score test, while its asymptotic properties are derived in Section \ref{testst}, and an estimation strategy and practical recommendations are detailed in Section \ref{estimst}. In Section \ref{simus}, we present numerical studies assessing the potential impact of ignoring the mean-variance relationship in RNA-seq data, and comparing our approach to state of the art methods of gene set analysis. In Section \ref{appli}, we apply tcgsaseq to the analysis of real data. Final remarks and comments are discussed in Section \ref{discuss}.

\section{Variance component score test for longitudinal gene set analysis}
\label{method}
\subsection{Problem setup} 
Consider $\br_i = (\br_{i1}\trans, ..., \br_{iP}\trans)\trans, \br_{ij} = (r_{ij1}, ..., r_{ijn_i})\trans$ a vector of sequence read counts for subject $i$, that have been mapped to each of $P$ genes measured at times $\bt_i = (t_{i1}, ..., t_{in_i})\trans$. 
Meanwhile $\bx_i = (x_{i1}, ..., x_{iq})\trans$ is a vector of baseline covariates describing experiment design conditions, all measured on individual $i$. So the full data considered for analysis are $n$ independent realizations from random vectors $\Vbb = \left\lbrace \bV_i = (\br_i\trans, \bt_i\trans, \bx_i\trans)\trans\right\rbrace_{i=1}^n$. Typically, the counts $\br_i$ are normalized in some way in a preprocessing step \citep{Hansen2012, SEQC2014}. We take $\by_i$ to be a normalized version of $\br_i$. For example, one standard normalization procedure accounts for the library size $\Lsc_{it} = \sum_{j=1}^P r_{ijt}$ by computing the log-counts per million as 
\begin{align}\label{eq:logcpm}
y_{ijt} = \log_2\left(10^6 \times \frac{0.5 + r_{ijt}}{1 + \Lsc_{it}}\right).
\end{align}

Based on these data, our interest is in identifying gene sets that have longitudinally changing expression patterns. The rest of this section will develop the proposed test statistic and its limiting distribution.

\subsection{The test statistic}
We are interested in testing for longitudinal changes in a pre-specified gene set, which for the purposes of illustration we take to be the first $p$ genes, $\by_i = (\by_{i1}, ..., \by_{ip})$. To develop a variance component score test statistic, we start from the following working model, which is a linear mixed effect model:
\begin{align}\label{mixed_model_vector}
&y_{ijt} = \alpha_{0j} + a_{0ij} + \bx_{i}\trans\balpha_j + \bphi_{it}\trans\bbeta_{j} + \bphi_{it}\trans\bxi_{ij} + \epsilon_{ijt}
\end{align}
and which can be rewritten as
\begin{align}\label{mixed_model}
&\by_i = \balpha_0 + \ba_{0i} + X_i\balpha + \Phi_i\bbeta + \Phi_i\bxi_i + \bvarepsilon_i\\
& \bxi_i = (\bxi_{i1}\trans, ..., \bxi_{ij}\trans)\trans \sim N(0, \Sigma_\xi), \quad \bvarepsilon_i \sim N(0, \Sigma_{\varepsilon i}), \quad \ba_{0i} \sim N(0, \Sigma_a)
\end{align}
where $\balpha_0$ is a $pn_i \times 1$ vector of gene-specific intercepts $\alpha_{0j}$, $\ba_{0i}$ is a $pn_i \times 1$ vector of random intercepts $a_{0ij}$, $X_i$ is a $pn_i \times pq$ block diagonal matrix with each block consisting of $n_i$ rows equal to the fixed effect baseline covariates $\bx_i$, $\balpha$ is a $pq \times 1$ vector of gene-specific fixed effects $\balpha_j$, $\Phi_i = \Phi(\bt_i)$ is a $pn_i \times pK$ block-diagonal matrix encoding the effect of time with each block having $t^{\text{th}}$ row $\bphi_{it} = \{\phi_1(t_{it}), ..., \phi_K(t_{it})\}$ for some set of $K$ basis functions $\{\phi_k(\cdot)\}$, $\bbeta$ is a $pK\times1$ vector of gene-specific fixed effects of time $\bbeta_j$, $\bxi_i$ is a $pK\times1$ vector of gene-specific random effects of time, $\Sigma_\xi = \cov(\bxi_i, \bxi_i)$, and $\Sigma_{\varepsilon i}$ is a $pn_i \times pn_i$ covariance matrix of measurement errors. Note that $\Sigma_{\varepsilon i}$ is indexed by $i$ and may depend on the mean of $\by_i$. We assume that $\ba_{0i} \perp \bvarepsilon_i$. It is important to note that, in practice, this model is unlikely to hold. Fortunately the testing procedure we propose is entirely robust to its misspecification.

We are interested in testing the null hypothesis of no longitudinal change in the normalized gene expression for any genes in the gene set \begin{align} \label{h0}
H_0: \text{The mean of $\by_i$ does not depend on $\bt_i$}
\end{align}
or, that is, $E(\by_i | \bt_i) = E(\by_i) \text{ and } E(\by_i | \bt_i, \bx_i) = E(\by_i | \bx_i)$.
Under the model \eqref{mixed_model}, the null hypothesis \eqref{h0} corresponds to the following null: $H_0: \bbeta = \bxi_i = \bzero$.
In section \ref{testst}, we demonstrate that the corresponding variance component score test can be written as 
\begin{align} \label{Q}
Q = \bq\trans\bq, \qquad \bq\trans = \nnhalf\sumin\by_{\mu i}\trans  \Sigma_i\inv \Phi_i\Sigma_\xi^{1/2}
\end{align}
where $\Sigma_i = \Sigma_a + \Sigma_{\varepsilon i}$, $\by_{\mu i} = \by_i - \balpha_0 - X_i\alpha$ and $\Sigma_\xi^{1/2}$ is the symmetric half matrix such that $\Sigma_\xi^{1/2}\Sigma_\xi^{1/2} = \Sigma_\xi$. We also show in the Supplementary Material that, given the dimension of $\bq$ ($pK$) is small relative to the number of individuals ($n$), the asymptotic distribution of the test statistic is a mixture of $\chi_1^2$ random variables, $
Q \longrightarrow \sum_{l =1}^{pK} a_l\chi^2_1,
$
where the mixing coefficients $\{a_l\}$ depend on the covariance of $\bq$. In the end, p-values may be computed by comparing the observed test statistic $Q$ to the distribution of $\sum_{l=1}^{pK}\ahat_{l}\chi_1^2$ where $\ahat_l$ is an estimate of $a_l$. Details are developed further in Section \ref{testst} and the Supplementary Material. When the entries of $\bq$ are correlated -- i.e. there is correlation between genes in the gene set \--- then the degrees of freedom for the test based on $Q$ ($\sum_{l=1}^{pK}a_l$) may be much lower than the degrees of freedom for a similar Wald or score test ($pK$), yielding more power to detect departures from the null hypothesis.

In practice, the parameters $\balpha_0$, $\balpha$, $\Sigma_i$, and $\Sigma_\xi$ are unknown and must be estimated. In particular, accounting for the mean-variance relationship encoded in $\Sigma_i$ is vital in RNA-seq data. Hence we estimate the test statistic as
\begin{align}\label{est_test}
\Qhat = \bqhat\trans\bqhat, \qquad \bqhat\trans = \nnhalf\sumin(\by_i - \balphahat_0 - X_i\balphahat)\trans  \Sigmahat_i\inv \Phi_i\Sigmahat_\xi^{1/2}
\end{align}
and we further demonstrate in section \ref{testst} that plugging in standard estimates of $\Sigma_\xi$, $\balpha_0$, and $\balpha$ and a nonparametric estimator of $\Sigma_i$ for the estimated test statistic still yields a similar asymptotic distribution:
\begin{align}\label{est_lim}
\Qhat \longrightarrow \sum_{l=1}^{pK}\atilde_{l}\chi_1^2
\end{align}
for mixing coefficients $\{\atilde_l\}$ given in the Supplementary Material.

The strength of our approach is that this simple limiting distribution holds even when the model \eqref{mixed_model} may be misspecified and despite the presence of the nonparametric estimator of $\Sigma_i$ in \eqref{est_test}. In this way, we may account very flexibly for the mean-variance relationship in $\by$ while maintaining a simple, powerful test statistic that does not require any particular model to hold. 

The convergence in \eqref{est_lim} relies on the central limit theorem. There may of course be situations where the central limit theorem fails to kick in, either when the number of genes in the gene set $p$ is quite large, or else when the number of individuals $n$ is small. We now discuss two modifications for these scenarios.

\subsubsection{Testing for homogeneous gene sets}
When the number of genes $p$ is large relative to $n$, the convergence of $\bq$ to a limiting normal distribution may be in doubt. In these cases, it might be better to begin from a working model that assumes that all genes share a common trajectory over time, as a useful approximation. To wit, we may reduce the number of parameters in the model by taking $X_i$ in model \eqref{mixed_model} to be a $pn_i \times q$ counterpart of $X_i$, and similarly $\balpha$ to be $q \times 1$, $\Phi_i$ to be $pn_i \times K$, and $\bxi_i$ to be $K\times1$ counterparts, respectively, of the original quantities.

The derivation of the test statistic in this case follows precisely the same lines, once dimensions are updated. This test will be powerful to detect alternatives where all genes in the set develop in a similar fashion longitudinally. However, this test should be expected to have low power to detect situations where some genes, for example, decline over time while other genes increase over time, as in heterogeneous gene sets.

\subsubsection{Using permutation}
When $n$ is very small (or alternatively  when $p$ is relatively large but the homogeneous strategy described above seems unwise), relying on the limiting distribution \eqref{est_test} may not be accurate. In these cases, permutation may be used to estimate the empirical distribution of $\Qhat$ under the null \eqref{h0}. To perform the permutations, we simply shuffle the time labels within each individual to get permuted observations for the $i$th individual $y_{ijt}^* = y_{ij\Isc_t}$ where $\{\Isc_1, ..., \Isc_{n_i}\}$ is a permutation of $\{1, ..., n_i\}$. Indeed, under the null, observations of a given gene $j$ for a given individual $i$ are exchangeable, regardless of sampling time. A large number $B$ of permutation-based test statistics $\mathfrak{Q} = \{Q^*_1, ..., Q^*_B\}$ can thus be generated where each $Q^*_b$ is computed using the $b$th set of permuted data. P-values may then be computed as:
\begin{align}
\phat = \dfrac{1}{B}\sum_{b=1}^B I\{\Qhat \leq Q^*_b\}
\end{align}

\section{Properties of the test statistic}
\label{testst}
In this section, we derive the test statistic and demonstrate its asymptotic distribution assuming all parameters known. We then take up the distribution of the test statistic when estimating all relevant parameters. 

\subsection{Test statistic derivation}

Under the working model \eqref{mixed_model}, $\displaystyle
\by_i | \ba_{0i}, \bxi_i, \bx_i, \bt_i \sim N(\balpha_0 + \ba_{0i} + X_i\balpha + \Phi_i\bbeta + \Phi_i\bxi_i, \Sigma_{\varepsilon i}).
$
Integrating over the random intercepts $\ba_{0i}$, we can rewrite the model as 
\begin{align}\label{int_model_norm}
\by_i | \bxi_i, \bx_i, \bt_i \sim N(\bmu_i + \Phi_i\btheta_i, \Sigma_{i})
\end{align}
where $\bmu_i = \balpha_0 + X_i\balpha$ denotes time-independent fixed effects, $\btheta_{i}=\left(\bbeta_{1}\trans + \bxi_{i1}\trans, \dots,\right.$ $\left.\bbeta_{p}\trans + \bxi_{ip}\trans\right)\trans$ denotes combined effects of time.

The test for no longitudinal changes in expression corresponds to the model-based hypothesis $H_0: \btheta_{i} = \bzero$. We write $\btheta_{i}$ as $\btheta_{i} = \eta\bnu_{i}$ and we consider the working assumption that the $\{\bnu_{i}\}$ are independently distributed such that $E(\bnu_{ij})=\bzero$ (under the null) and
$\var(\bnu_{ij}) = \Sigma_\xi$.
Under this working assumption, $H_0: \btheta_{i} =\bzero$ is equivalent to $H_0: \eta = 0$. To obtain the variance component test statistic, rewrite the model as: $\by_{\mu i} =\Phi_i\btheta_i + \bvarepsilon_{i}$ for centered outcome $\by_{\mu i} = \by_{i} - \bmu_{i}$. Then $\left.\by_{\mu i}\right| \bnu_i,\sim N\left(\Phi_i\eta\bnu_i, \Sigma_i\right)$ and the log-likelihood for $\by_{\mu i}$ can be written as 
\begin{align}
\log\Lsc(\eta) =
-\frac{1}{2} \sumin\left\lbrace\log\left|\Sigma_i\right| + 
\left(\by_{\mu i} - \eta\Phi_i\bnu_i\right)\trans \Sigma_i\inv \left(\by_{\mu i}  - \eta\Phi_i\bnu_i\right)\right\rbrace .
\end{align}
Because the target of inference is $\eta$, we marginalize over the nuisance parameter $\bnu$ conditional on the observed data to obtain
$\Lsc^*(\eta) = E\{\Lsc(\eta) | \mathbb{V}\}$ where the expectation is taken over the distribution of $\bnu$. We follow the argument in \citet{Commenges1995} and note that the score at the null value is 0: $\lim_{\eta \to 0}\partial \log\Lsc^*(\eta) / \partial \eta = E\left(\sumin \by_{\mu i}\trans \Sigma_i\inv \eta\bnu_i\trans\bphi_i\ \mid \Vbb\right) = 0$. So we instead consider the score with respect to $\eta^2, \lim_{\eta \to 0} \partial \log \Lsc^*(\eta) / \partial (\eta^2)$, and 
\begin{align}
&E\left\{\left.\frac{\partial \log\Lsc(\eta)}{\partial \eta}\right|_{\eta = 0} \mid \Vbb\right\}^2 + \text{constant} + o_p(1)\\
&= E\left(\eta\sumin \by_{\mu i}\trans \Sigma_i\inv \Phi_i\bnu_i \mid \Vbb\right)^2  + \text{constant} + o_p(1)\\
&= \left(\sumin\by_{\mu i}\trans  \Sigma_i\inv \Phi_i\right)\Sigma_\xi\left(\sumin\by_{\mu i}\trans  \Sigma_i\inv \Phi_i\right)\trans  + \text{constant} + o_p(1)
\end{align}
Thus, after normalizing by $n\inv$, the variance component score test statistic can be written $Q = \bq\trans\bq$, as described in \eqref{Q}. If the dimension of $\bq$ is small relative to the sample size, $\bq$ is asymptotically normal by the central limit theorem, which means that the limiting distribution of $Q$ is a mixture of $\chi^2_1$s (see Supplementary Material for more details).

\subsection{Estimated test statistic limiting distribution}

In practice, estimates need to be supplied for many of the quantities in the test statistic to be estimated in \eqref{est_test}. Following the argument for the limiting distribution of $Q$, $\Qhat$ has a limiting distribution of a mixture of $\chi^2_1$s so long as $\bqhat$ has a limiting normal distribution. To establish this, we define $\bqtilde = \nnhalf\sumin (\by_i - \bmuhat_i)\trans\Tsc_i$ to be a version of $\bqhat$ where $\Sigmahat_i\inv\Phi_i\Sigmahat_\xi^{1/2}$ is replaced with its limit, $\Tsc_i = \lim_{n \rightarrow \infty} \Sigmahat_i\inv\Phi_i\Sigmahat_\xi^{1/2}$. 
The quantities $\bqhat$ and $\bqtilde$ are asymptotically equivalent, and the central limit theorem ensurea that $\bqtilde$ has a limiting normal distribution. This further suggests that the asymptotic distribution of $\Qhat$ is a mixture of $\chi_1^2$s (see Supplementary Material for details).

\section{Estimation}
\label{estimst}

The test statistic \eqref{est_test} contains the estimated quantities $\balphahat_0$, $\balphahat$, $\Sigmahat_{\xi}$, and $\Sigmahat_i$. In this section, we take up practical issues involved in the estimation of these quantities. 

\subsection{Estimating model parameters}
In this section, we will discuss estimation of $(\balphahat_0, \balphahat, \Sigmahat_{\xi})$, assuming that we have a consistent estimator of $\Sigmahat_i$ in hand (we leave discussion of estimating $\Sigmahat_i$ to the following section). A natural way to estimate $(\balphahat_0, \balphahat, \Sigmahat_{\xi})$, taking into account the heteroscedasticity in $\by$, is to fit a weighted mixed effects model corresponding to \begin{align}
y_{ijt} = \alpha_{0j} + \bx_{i}\trans\balpha_j + \bphi_{it}\trans\bbeta_{j} + \bphi_{it}\trans\bxi_{ij} + \epsilontilde_{ijt},\label{integrated_model}
\end{align}
where the random intercepts $a_{0ij}$ have been integrated out. The weights are taken to be $\bw_i = \text{diag}(\Sigmahat_i)\inv$.  This model corresponds to the model \eqref{int_model_norm}. 

In practice, fitting the full mixed effects model for each of many gene sets may be computationally demanding. In these cases, the simpler fixed effects model  \begin{align}\label{simple_model}
y_{ijt} = \alpha_{0j} + \bx_{i}\trans\balpha_j + \bphi_{it}\trans\bbeta_j + \epsilontilde_{ijt}
\end{align}
may be fit to estimate $\balpha_0$ and $\balpha$. $\Sigma_{\xi}$ may not be estimated directly from \eqref{simple_model}. In this case, it may be specified using a working estimate or taken to be the identity matrix.

\subsection{Estimating the mean-variance relationship}\label{nuisance}

One key feature of our approach is to estimate $\cov(\by_i | \bx_i, \bt_i, \bxi_i) = \Sigma_{i}$ in such a way as to account for the heteroscedasticity in $\by$. It is important to note that accurate estimation of $\Sigma_i$ will likely increase power but will not affect the validity or type I error of the testing procedure.

To approximate the mean-variance relationship in $\by$, we use information from all $P$ genes. We assume that the diagonal elements of $\Sigma_i$, which we will denote $\bv = \{v_{ijt}\} = \{\var(y_{ijt} | \bx_i, \bt_i, \bxi_i)\}$, may be modeled as a function of their means $\bm = \{m_{ijt}\} = \{E(y_{ijt} | \bx_i, \bt_i, \bxi_i)\}$. Specifically,
$v_{ijt} = \omega\{m_{ijt}\} + e_{ijt}$
for some unknown function $\omega(\cdot)$ and errors which follow the moment conditions $E(e_{ijt}) = 0, \var(e_{ijt}) = \tau^2, \tau > 0$.



We follow \citet{Law2014} in using local linear regression in estimating $\omega(\cdot)$. We will first write out the form of the estimator if all parameters were known \begin{align}
\omegatilde_n(x) = \sum_{ijt} \tilde{\ell}_{ijt}(x)v_{ijt}, \quad\tilde{\ell}_{ijt}(x) = \frac{\btilde_{ijt}(x)}{\sum_{ijt} \btilde_{ijt}(x)}\\ \btilde_{ijt}(x) = K\{(m_{ijt} - x)/h\}\left\{\Stilde_{n2}(x) - (m_{ijt} - x)\Stilde_{n1}(x)\right\}
\end{align}
where $\Stilde_{nd}(x) = \sum_{ijt} K\{(m_{ijt} - x)/h\}(m_{ijt} - x)^d$ for some kernel function $K(\cdot)$ and bandwidth $h > 0$.

The means $\bm$ and variances $\bv$ could in principle be estimated using some parametric model like \eqref{integrated_model} for all $P$ genes. Let $(\mhat_{ijt}, \vhat_{ijt})$ denote the estimated mean and variance for $y_{ijt}$. Then the mean-variance relationship may be estimated as
\begin{align}\label{omegahat}
\omegahat_n(x) = \left.\omegatilde_n(x)\right|_{m_{ijt} = \mhat_{ijt}, v_{ijt} = \vhat_{ijt}}
\end{align}

As with any smoothing procedure, choice of the bandwidth $h$ is paramount in producing a reliable estimator. Standard cross-validation techniques may be used to select $h$ in practice. Then the diagonal entries of $\Sigma_i$ can be estimated as $\omegahat_n(\bmhat_i)$.

\paragraph{Practical considerations}

Under the model \eqref{integrated_model}, we could take $\mhat_{ijt} = \alphahat_{0j} + \bx_{it}\trans\balphahat_j + \bphi_{it}\trans\bbetahat_j + \bphi_{it}\trans\bxihat_{ij}$ and $\vhat_{ijt} = (y_{ijt} - \alphahat_{0j} - \bx_{it}\trans\balphahat_j - \bphi_{it}\trans\bbetahat_j - \bphi_{it}\trans\bxihat_{ij})^2$. However, because $P \gg n$, fitting \eqref{integrated_model} for such a large number of genes is not practical. Instead, the simpler model \eqref{simple_model} could be used. Then $\mhat_{ijt}$ could be taken to be $\alphahat_{0j} + \bx_{it}\trans\balphahat_j + \bphi_{it}\trans\bbetahat_j$ and $\vhat_{ijt}$ to be $(y_{ijt} - \alphahat_{0j} - \bx_{it}\trans\balphahat_j - \bphi_{it}\trans\bbetahat_j)^2$. A similar approach was used to model heteroscedasticity in the context of linear regression in \citet{Carroll1982}.

On the other hand, fitting a local linear regression on all of $\bm$ and $\bv$ -- a total of $P\sumin n_i$ observations -- may be computationally difficult. In order to reduce the number of points used in the nonparametric fit \eqref{omegahat}, one could follow \citet{Law2014} and model the mean-variance relationship at the gene level \begin{align}
v_j = \omega(m_j) + e_j.\label{genelevel}
\end{align}
The gene-level mean may be estimated as $\mhat_j = n\inv\sumin n_i\inv\sum_{t=1}^{n_i}\alphahat_{0j} + \bx_{it}\trans\balphahat_j + \bphi_{it}\trans\bbetahat_j$ and the gene-level variance as $\vhat_j = n\inv\sumin n_i\inv\sum_{t=1}^{n_i}(y_{ijt} - \alphahat_{0j} - \bx_{it}\trans\balphahat_j - \bphi_{it}\trans\bbetahat_j)^2$.

\section{Numerical study}
\label{simus}

In this numerical study section, we illustrate both the importance of accounting for RNA-seq data's heteroscedasticity and the superiority of tcgsaseq in terms of statistical power and robustness. We have performed simulations under two different settings. The first one demonstrates the good behavior of the asymptotic test under various scenarios in synthetic data. The second one focuses on a very realistic situation where sample size is small and gene counts are generated from a distribution of real observed RNA-seq data. Throughout, for the purposes of testing, we take $\Sigma_a$ and $\Sigma_\xi$ to be the identity for simplicity. In practice, more precise estimation of $\Sigma_a$ and $\Sigma_\xi$ would serve to increase power. 

\subsection{Synthetic data}

In this subsection, we use synthetic data to illustrate the behavior of our estimator when the distribution of the data is known. We first look at the importance of accounting for the mean-variance relationship in $\Sigma_i$ in tcgsaseq, and then we will compare tcgsaseq to other competing methods, including ROAST, edgeR-ROAST, and DESeq2. In order to adapt DESeq2 to self-contained gene set testing, we use the minimum p-value test to adequately combine univariate p-values for testing a whole gene set while taking into account gene correlation \citep{Moskvina2008a, Lin2011Kernel} and we refer to it as DESeq2-min test. We demonstrate that the tcgsaseq testing procedure is robust to even heavy misspecification of the mean model and the mean-variance relationship, while ROAST, edgeR-ROAST and DESeq2-min test may suffer from extreme lack of power compared to tcgsaseq or inflate type I errors under misspecification.

\subsubsection{Mean-variance relationship}
To illustrate the importance of estimating the mean-variance relationship, we generated data under the model $\displaystyle y_{ijt} = \log\left\{\frac{(\mu_{ijt}+0.5) 10^6}{\sum_i \mu_{ijt} +1}\right\}$ where
	\begin{align}\label{synth_data_model}
    \mu_{ijt} = \eta_{ijt}\sum_{i, t} \eta_{ijt}/P + b_{ij}t_{it} + \beta t_{it}, \qquad
    \eta_{ijt} = a_{ij} + \alpha_{j} + \sum_{k=1}^3x_{itk} + \epsilon_{ijt},
    \end{align}
with $\epsilon_{ijt} \sim N(0, \alpha_j^2)$, $\alpha_{j} \sim \text{exp}(100)$, $a_{ij} \sim N(0, \alpha_j^2/100)$, $x_{itk} \sim N(100, 2500)$, $t_{it} \sim U(0,1)$, $b_{ij} \sim N(0, 1)$. We set $n = 200, p = 10, P = 1000, n_i = 3,$ and $\beta$ was allowed to be set to 6 different values within a range from $0$ to $2$: $(0, 0.5, 0.75, 1, 1.25, 1.5, 2)$. We considered three methods of accounting for the mean-variance relationship: (i) gene-level estimates using equation \eqref{genelevel}; (ii) voom-type estimation \citet{Law2014}; (iii) specifying $\Sigma_i$ to be the identity. Throughout this section we test at the $0.05$ level. The model \eqref{simple_model} was used to generate the test statistic and adjust for covariates, and the gene-level equation \eqref{genelevel} was used to estimate the mean-variance relationship. Note that the mean model \eqref{simple_model} is misspecified. 

When applying our variance component test, type I error was at the nominal level using either strategy (i) or (iii), but was deflated to 0.035 when using strategy (ii). On the other hand, ROAST, which similarly relies on estimating the mean-variance relationship, inflated type I error to 0.56 when using (ii), and had adequate type I errors using the other weighting strategies. Such results indicate that ROAST may drastically inflate type I error rates in some cases, and that using (ii), the voom-type estimator, to account for heteroscedasticity led to worse performance in general. 

\begin{figure}[!h]
\centering
\begin{tabular}{cc}
\includegraphics[ width = 0.55\textwidth, keepaspectratio]{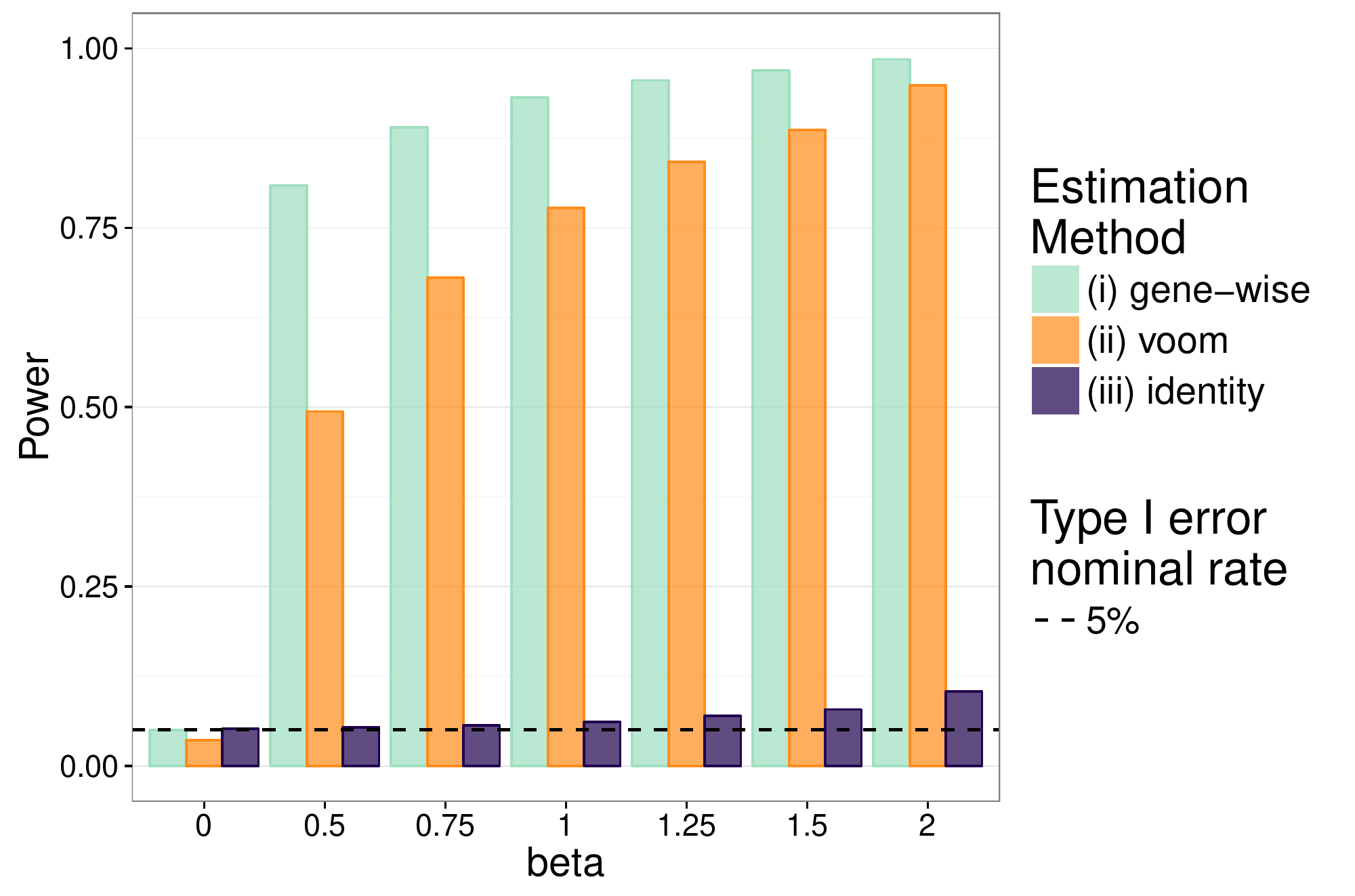}
\end{tabular}
    \caption{Power evaluation in synthetic data according to how heteroscedasticity is accounted for, based on 1,000 simulations.}
    \label{synth_data}
\end{figure}

Figure \ref{synth_data} demonstrates the effect of accounting for heteroscedasticity under the alternative hypothesis. The more accurate modeling of the mean-variance relationship with method (i) yields large power gains over method (ii). Conversely, by drastically mis-modeling the mean-variance relationship, the naive method (iii) yields almost no power for detecting longitudinal changes. 

\subsection{Comparison to competing methods}
In this subsection, we compare tcgsaseq to ROAST, edgeR-ROAST, and DESeq2-min test using a highly misspecified model and using a negative binomial model. For the misspecified model, we again generate data under model \eqref{synth_data_model} with one modification to ensure that the data are positive integers $y_{ijt} =$\\$\max\left\{\left\lceil\log\left(\frac{(\mu_{ijt}+0.5) 10^6}{\sum_i \mu_{ijt} +1}\right)\right\rceil, 1\right\}$. We set $n = 50, 100, 150$ and $n_i = 5$, and $\beta$ took values between -2 and 2. Because DESeq2 is computationally intractable at higher sample sizes with large $P$, we set the total number of genes $P$ to be 100. We used model \eqref{genelevel} to compute the mean-variance relationship for both tcgsaseq and ROAST. Gene-based estimates were used for dispersions in DESeq2, and likelihood ratio tests were used to produce test statistics for both edgeR and DESeq2. 

\begin{figure}[!h]
\centering
\begin{tabular}{cc}
\includegraphics[ width = \textwidth, keepaspectratio]{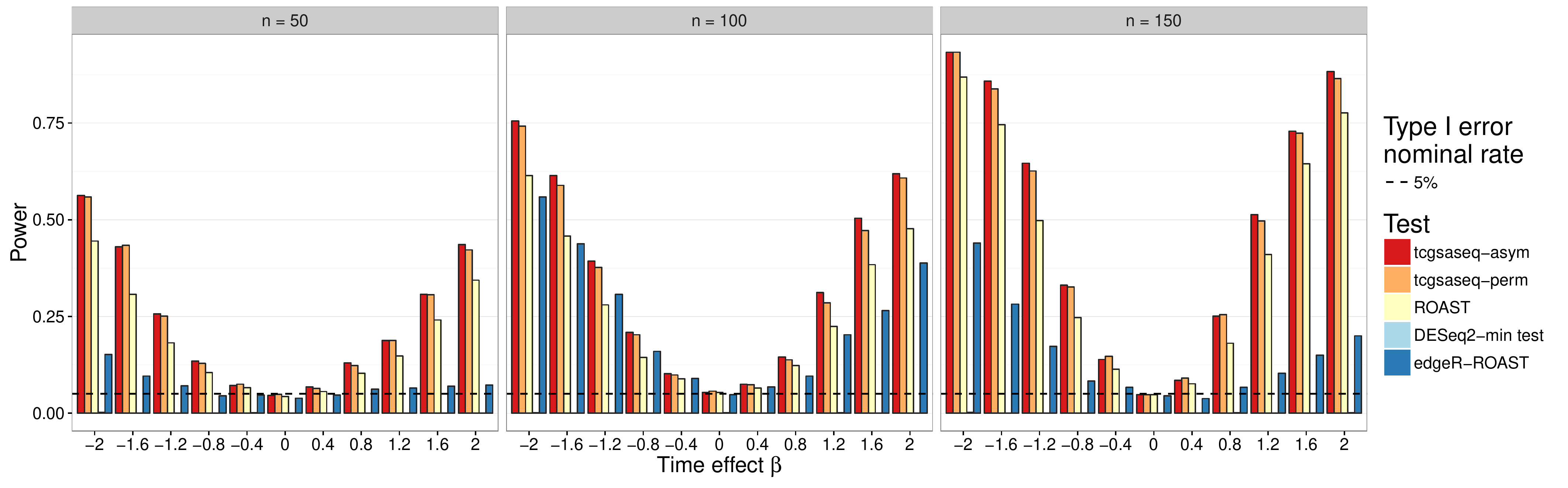}
\end{tabular}
    \caption{Power evaluation in synthetic data comparing tcgsaseq, ROAST, edgeR-ROAST, and DESeq2-min test, based on 1,000 simulations.}
    \label{synth_power}
\end{figure}

The results are depicted in Figure \ref{synth_power}. We see that tcgsaseq has the highest power at all sample sizes and at all values of $\beta$. Though the model is highly misspecified for all methods, the negative-binomial-based methods edgeR-ROAST and DESeq2-min test suffer greatly in terms of power, in particular DESeq2-min test. In Figure \ref{synth_power}, we show ROAST using the mean-variance relationship estimated from model \eqref{genelevel}, but it's important to note that if the voom-type strategy were used, the type I error for ROAST rises to more than 0.1 at all sample sizes. 

\begin{figure}[!h]
\centering
\begin{tabular}{cc}
\includegraphics[ width = \textwidth, keepaspectratio]{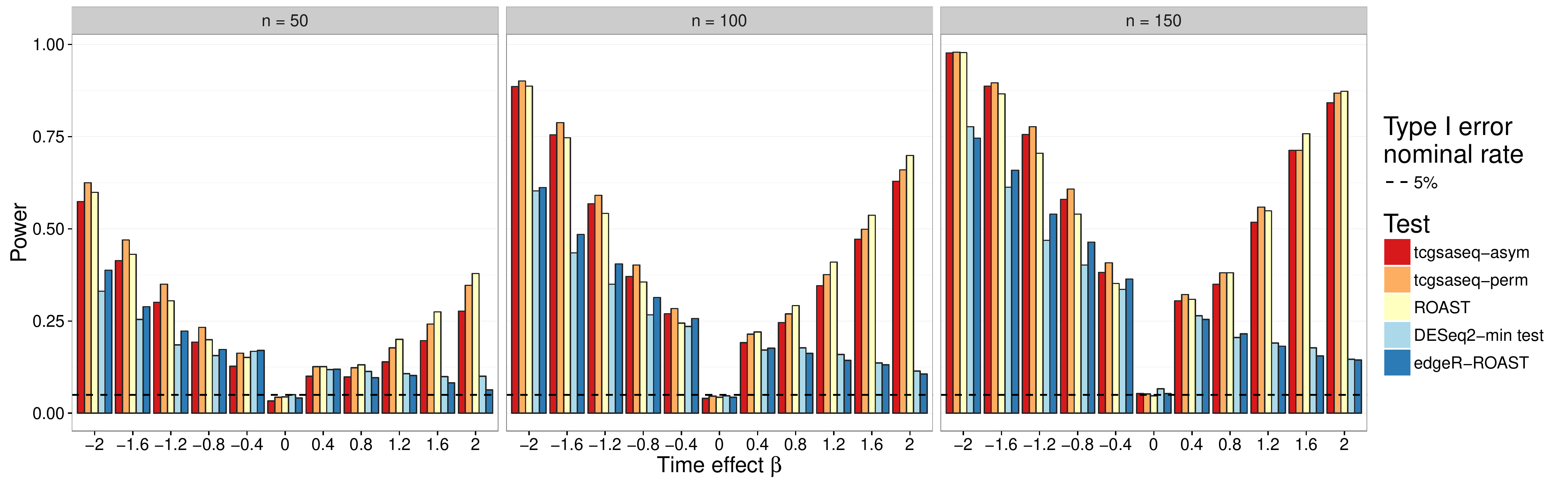}
\end{tabular}
    \caption{Power evaluation in negative binomial data comparing tcgsaseq, ROAST, edgeR-ROAST, and DESeq2-min test, based on 1,000 simulations.}
    \label{nb_power}
\end{figure}

Secondly, we generated data under the negative binomial model, a distributional assumption that edgeR and DESeq2 make. The mean of the negative binomial was specified as 
\begin{align}\label{nb_model}
\mu_{ijt} = \max\{1001 + a_{0i} + x_i + (b_{i} + \beta + \beta_j)t_{it}x_i, 0\}
\end{align}
where $a_{0i} \sim N(0,1), x_i \sim N(\mu_{xi}, 1), \mu_{xi} \sim$  exponential(1/10), $b_i \sim N(0,1)$, $\beta_j \sim N(0,1)$, and the negative binomial dispersion parameter was set to 1. We again let $\beta$ vary between -2 and 2 and considered sample sizes of $n = 50, 100, 150$. In this setting, the default procedure to estimate dispersions was used for DESeq2. The results depicted in Figure \ref{nb_power} show that tcgsaseq and ROAST outperform edgeR-ROAST and DESeq2-min test despite the fact that the distributional assumptions of edgeR and DESeq2 are true. The performance of tcgsaseq and ROAST is in general comparable
. Furthermore, it is important to highlight that DESeq2-min test does not control the type I error, and the type I error gets worse as sample size increases. 

\vspace{10cm}


\subsection{Realistic small samples simulations}

In this subsection, we simulated another dataset to illustrate the good behavior of the tcgsaseq permutation test in the realistic setting of a small sample size. We generated data for 6 individuals each measured at 3 time points according to the scenario described in \cite{Law2014}, using the script provided in their Supplementary Material:
\begin{align}
m_{ij} = 0.2 + 1/\sqrt{\mu_{ij} + \beta t_{it} }\qquad\text{ and }\qquad y_{ijt} = \log\left(\frac{(a_{ij}+0.5) 10^6}{\sum_j a_{ij} +1}\right)
\end{align}
where $t_{it} \in \{1, 2, 3\}$, $a_{ij}\sim \text{Pois}(\lambda_{ij})$, $\lambda_{ij}\sim \text{Gamma}(\frac{\kappa_{ij}}{40 m_{ij}}, \frac{{40 m_{ij}}}{\kappa_{ij} \mu_{ij}})$, $\kappa_{ij}\sim\chi^2(40)$, and $\mu_{ij}$ follows an empirical baseline distribution derived from real RNA-seq counts data, provided in supplementary material of \cite{Law2014}. This simulation scheme allows us to generate data that realistically resemble real RNA-seq count data. We set $n=6$ with $n_i=3$, and gene sets were constructed such that $p \in (30, \dots, 400)$ and for every gene pair $(j, j')$ in the set $\text{cor}(\bmu_{j}, \bmu_{j'})>0.8$. $\beta$ was allowed to be set to 8 different values within a range from $0$ to $0.5$: $(0, 0.01, 0.05, 0.1, 0.2, 0.3, 0.4, 0.5)$.

\begin{figure}[!h]
\centering
	\includegraphics[width=0.95\textwidth]{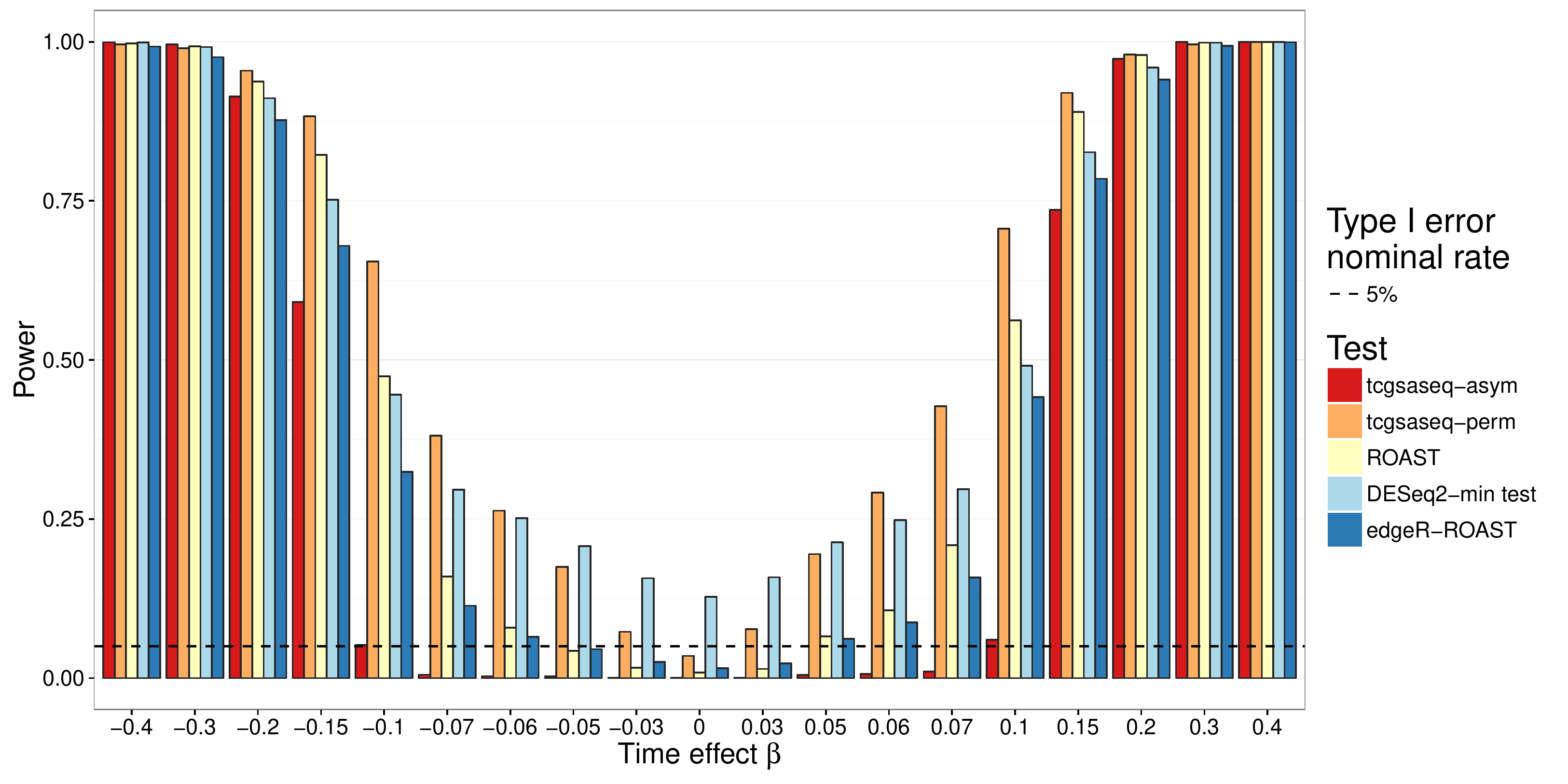}
    \caption{Power evaluation in realistically simulated data with a small sample size, based on 500 simulations.}
    \label{power_voomlike}
\end{figure}

As shown in Figure \ref{power_voomlike}, under this setting, edgeR-ROAST, ROAST and the proposed permutation test all control the type I error at nominal rate. However DESeq2-min test fails to control the type I error, while it is highly deflated for the asymptotic version of tcgsaseq (0.001) which is not surprising given that with such a small sample size it is unlikely the central limit theorem would have come into effect yet. Nonetheless, we observe a steady and consistent increase of power for our method using the permutation test over the compared state of the art approaches. The deviation from the negative binomial distribution hypothesis (which both edgeR and DESeq2 rely upon), as well as the univariate gene-wise step first needed, explain those poor performances compared to tcgsaseq. Similar results are obtained when adding a random effect of time simulating heterogeneous gene sets (see Supplementary Material). In addition, as gene sets were constructed with correlated genes, these results also indicates that tcgsaseq is robust to and can take advantage of inter-gene correlation.

\section{Application to two real datasets}
\label{appli}

In this section we present analyses of two real datasets. In both analyses, $\Sigma_a$ and $\Sigma_\xi$ were again taken as the identity while $\Sigma_i$ was estimated at the gene level using equation \eqref{genelevel} for tcgsaseq. The ROAST method was applied in combination with precision weights estimated through the voom approach. Given the sample sizes available in both studies, we used the tcgsaseq permutation test. In both case, only transformed data were available and not original counts data, preventing us from applying either edgeR-ROAST or DESeq2-min test.

\subsection{Longitudinal RNA-seq measurements in successful kidney transplant patients}

We analyzed a RNA-seq dataset from \cite{Dorr2015}, in which gene expression was measured in the peripheral blood mononuclear cells of 32 kidney transplant patients. Gene expression was measured at 4 time points: before transplantation, 1 week after transplantation, 3 months after transplantation and 6 month after transplantation. The patients had had no graft rejection at the time of each sample.

We investigated custom gene sets targeted specifically towards kidney transplant. Those kidney-oriented gene sets were derived by the Alberta Transplant Applied Genomics Center from specific pathogenesis-based transcripts \citep{Halloran2010, Sellares2013, Broin2014}, and their definition is available at \url{http://atagc.med.ualberta.ca/Research/GeneLists}. We tested for a linear change in gene expression over time while adjusting for patient's age and gender. 

\begin{figure}[!h]
\centering
	\includegraphics[width=0.8\textwidth]{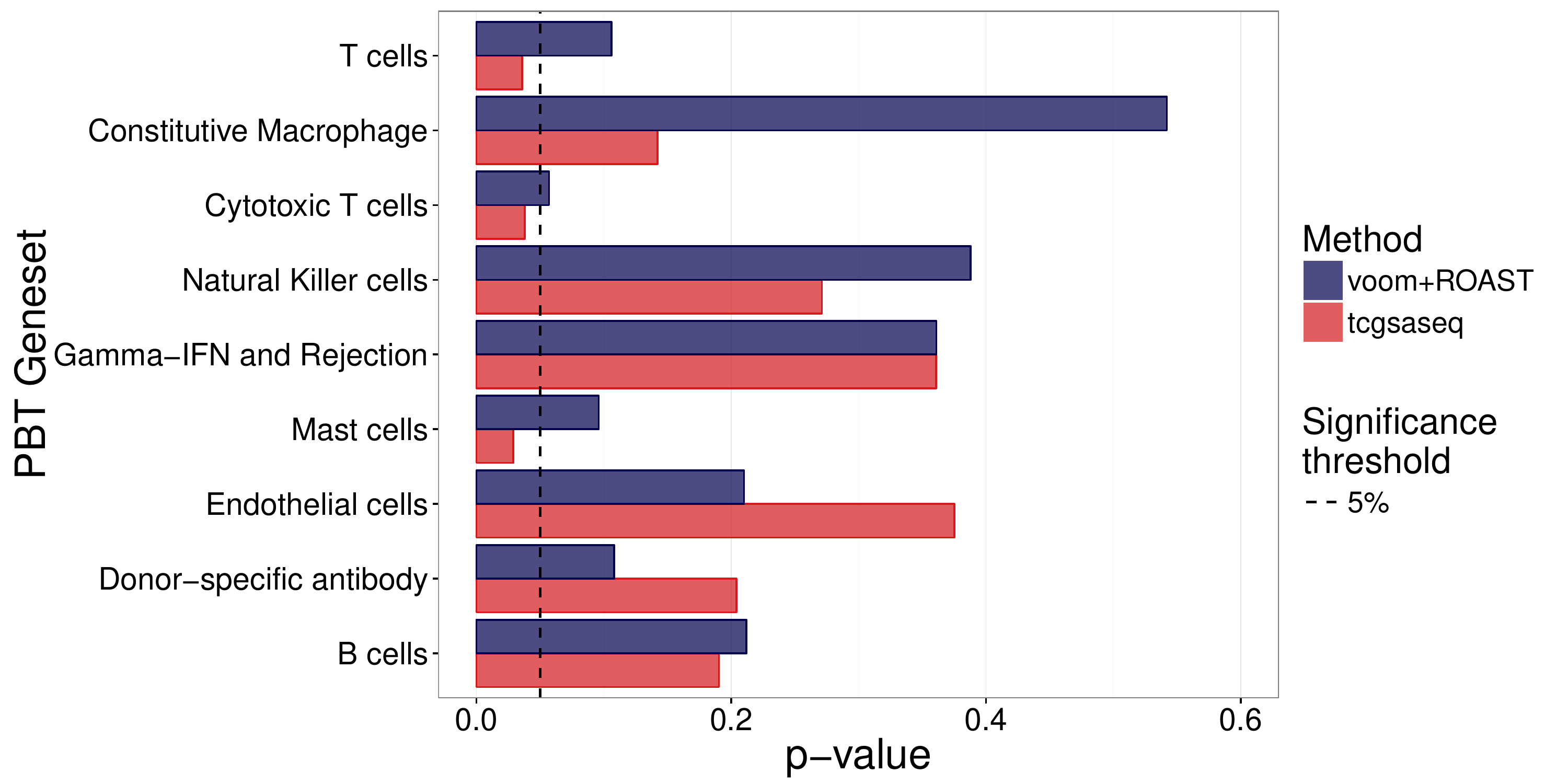}
    \caption{P-values from testing the 9 kidney oriented gene sets investigated}
    \label{kidneyPBT}
\end{figure}

Figure \ref{kidneyPBT} shows the p-values for those 9 gene sets for both tcgsaseq and ROAST. At a 5\% threshold, our approach tcgsaseq identifies three significant gene sets while the combination of voom and ROAST identifies none. Among those significant gene sets, two relate to T-cell gene expression, corroborating the original results from \cite{Dorr2015}. The gene set annotated as "T-cells" notably includes the gene CD3D, previously highlighted in \cite{Dorr2015}. On the contrary, gene sets related to transplant rejection such as "Donor-specific antibody" or "Gamma-IFN and Rejection" have a much higher p-value here, which is what one would expect as these data include only successful transplant patients. Besides, tcgsaseq also detects a significant change in expression for the gene set related to Mast cells, which have recently been highlighted as playing an ambiguous role in kidney transplant \citep{Papadimitriou2013a}. Mast cells have been linked both to peripheral tolerance \citep{DeVries2009}, as well as to late graft loss \cite{Jevnikar2008}. These results both reinforce and broaden the original findings from \cite{Dorr2015}.

\subsection{Time-course RNA-seq comparative study of \textit{Arabidopsis arenosa} physiology}

In a recent experiment, \cite{Baduel2016} measured time-course gene expression of the plant \textit{Arabidopsis arenosa} through RNA-seq. They sampled 48 plants across 13 weeks at four different time points. \cite{Baduel2016} were especially interested in the difference between two populations of \textit{Arabidopsis arenosa}, respectively denoted KA and TBG, that have adopted different flowering strategies. In addition, half of the plants were exposed to cold and short day photoperiods (vernalization) between week 4 and week 10 in order to study the corresponding effect on flowering in both populations. Different plant siblings were sampled at each time points in order to avoid the important stress effect of leaf removal on the plants.

Dealing with this complex experiment design, we used tcgsaseq to address two separate biological questions: i) which gene sets have a different activation between the two populations, adjusted for the plant age and the cold exposure, ii) which gene sets have a different activation due to cold exposure, adjusted for population differences and plant age. Using two different modeling strategies, we found that one data-driven gene set constructed by \cite{Baduel2016} from the top 1\% differentially expressed genes between the two populations was significant at a 5\% threshold for i) and for ii). On the contrary two other data-driven gene sets, again constructed by \cite{Baduel2016} 
and identified as the respective population-specific responses to cold exposure, were both significant at this 5\% threshold for ii) but not for i). In addition, two gene sets from Gene Ontology associated with salt and cold response pathways, respectively, were also investigated and found significant at the 5\% level for ii) but not for i). In comparison, ROAST gave similar results for i), but lacked power for ii) identifying only 1 out of 4 significant gene sets. This analysis corroborates the results obtained by \cite{Baduel2016}, and it further illustrates the good behavior of tcgsaseq in complex time-course RNA-seq studies.

\section{Discussion}
\label{discuss}

The proposed method detailed in this article constitutes an innovative and flexible approach for performing gene set analysis of longitudinal RNA-seq gene expression measurements.
The approach relies on a principled variance component score test that accounts for the intrinsic heteroscedasticity of RNA-seq data, and for which we derive a simple limiting distribution without requiring any particular model to hold. As illustrated in the previous sections, the good performance of the method when applied to various datasets constitutes a major strength of the method.

Our numerical study shows the importance of taking into account heteroscedasticity when analyzing RNA-seq data. We also demonstrate the robustness of our testing procedure to model misspecification. When comparing our proposed approach to ROAST, a state-of-the-art GSA method suitable for longitudinal studies, we illustrate superior statistical power. In addition, ROAST relies on the correct specification of a linear model and normality of errors, while our testing procedure does not make any modeling assumptions. Furthermore, despite their growing popularity, RNA-seq studies may yet face small sample sizes. In order to deal with this issue, we provide a permutation alternative to our asymptotic test. This alternative test also exhibits very good properties in our simulation studies. 

We show that widely used methods such as ROAST, edgeR or DESeq2 are unfit for time course gene set analysis. In realistic settings, those methods can fail to control the Type I error. And even in favorable settings, the proposed method outperforms all of them in terms of statistical power.

Of particular biological interest in longitudinal studies, the proposed approach can test for both homogeneous and heterogeneous gene sets simultaneously. This is especially relevant for gene sets constructed from biological pathways. It also results in a power gain compared to GSA methods sensitive only to homogeneous sets. In addition, the proposed solution for accounting for heteroscedasticity can also deal with non-count data.
 
As demonstrated by the simulation study, the proposed approach is also robust to inter-gene correlation within tested gene sets. While in our simulations we do not estimate inter-gene correlations, one could account for them more formally by following \cite{Wang2009} and estimating a working correlation matrix with the residuals from an initial gene-wise modeling of $\by$. The resulting estimates could be incorporated into the structure of $\Sigma_i$ to increase power.
 
Finally, in this paper we focus primarily on longitudinal measurements of RNA-seq. However, our approach directly applies to virtually all RNA-seq study designs, including traditional case-control and more complex studies. Our approach allows researchers to incorporate the natural heteroscedasticity in the data into a powerful test statistic that makes no modeling assumptions. Of note, the inclusion of time-varying covariates would require further assumptions concerning the model to be made to ensure the limiting distribution of the test statistic. Evaluating tcgsaseq's performance in a broader array of studies is an area for future research.

\section*{Software}
\label{soft}

Software in the form of R code is available on the Comprehensive R Archive Network as an R package \texttt{tcgsaseq}.

\section*{Acknowledgments}
This work was supported by the National Institutes of Health (NIH) [U54 HG007963 to B.P.H.]. The authors express their deepest gratitude to Professor Tianxi Cai, Harvard University, for her help and support in this work. They also thank Pierre Baduel for his help in analyzing the plant physiology data. \medskip

\noindent This article has been accepted for publication in \textit{Biostatistics} Published by Oxford University Press.\medskip

\noindent {\it Conflict of Interest}: None declared.

\section*{Supplementary Material}
\renewcommand{\thesection}{S}
\label{supMat}

\subsection{Properties of the test statistic}
In this section, we establish the limiting distribution of the quantity $Q$ and its estimated counterpart $\Qhat$. 

\subsubsection{Limiting distribution of $Q$}
Recall that $Q = \bq\trans\bq$. Let $\Gamma = \cov(\bq)$. Then 
\begin{align}
Q &= \bq\trans\Gamma^{-1/2}\Gamma\Gamma^{-1/2}\bq\\
&= \bu\trans UAU\trans \bu + o_p(1)
\end{align}
where $U$ is an orthonormal matrix of the eigenvectors of $\Gamma$, $A$ is a diagonal matrix of the eigenvalues of $\Gamma$, and $\bu = \Gamma^{-1/2}\bq$ is asymptotically standard multivariate normal by the central limit theorem. Noting that, because $U$ is orthonormal, $U\trans\bu$ is also asymptotically standard normal, $\bu\trans UAU\trans \bu = \sum_{k=1}^{pK} a_k (u^*_k)^2$ where $u^*_k$ is an element of the asymptotically standard normal $U\trans\bu$ and $a_k$ an eigenvalue of $\Gamma$. It finally follows that $Q \sim \sum_{l =1}^{pK} a_l\chi^2_1$.

\subsubsection{Limiting distribution of $\widehat{Q}$}
Recall that $\Qhat = \bqhat\trans\bqhat$. We first show that $\bqhat$ and $\bqtilde$ are asymptotically equivalent.
\begin{align}
\bqhat\trans - \bqtilde\trans &= \nnhalf\sumin (\by_i - \bmuhat_i)\trans\Sigmahat_i\inv\Phi_i\Sigmahat_\xi^{1/2} - \nnhalf\sumin (\by_i - \bmuhat_i)\trans\Tsc_i\\
&=\nnhalf\sumin (\by_i - \bmuhat_i)\trans(\Tschat(\bxtilde_i) - \Tsc(\bxtilde_i))
\end{align}
for the processes $\Tsc(\bxtilde_i) = \Sigma(\bxtilde_i)\inv\Phi(\bxtilde_i)\Sigma_\xi^{1/2} = \Sigma_i\inv\Phi_i\Sigma_\xi^{1/2}$ and $\Tschat(\bxtilde_i) =$\\$\Sigmahat(\bxtilde_i)\inv\Phi(\bxtilde_i)\Sigmahat_\xi^{1/2} =$ $\Sigmahat_i\inv\Phi_i\Sigmahat_\xi^{1/2}$ as a function of $\bxtilde_i = (\bx_i, \bt_i)$ where $\Sigma(\bxtilde_i) = \Sigma_i, \Phi(\bxtilde_i) = \Phi_i$ and $\Sigmahat(\bxtilde_i) = \Sigmahat_i$. Here we use the fact that $\Sigma_i$ only depends on $\bxtilde_i$. 

We can thus write
\begin{align}
\bqhat - \bqtilde &=\nnhalf\sumin\int (\by_i - \bmuhat_i)\trans(\Tschat(\bxtilde) - \Tsc(\bxtilde))d\left\{I\{\bxtilde_i \leq \bxtilde\}\right\}\\
&=\int (\Tschat(\bxtilde) - \Tsc(\bxtilde))\trans d\left\{\nnhalf\sumin(\by_i - \bmuhat_i) I\{\bxtilde_i \leq \bxtilde\}\right\}
\end{align}
Now since $E(\by_i - \bmuhat_i | \bxtilde_i) = 0$ under 
(2.4),
$\nnhalf\sumin(\by_i - \bmuhat_i) I\{\bxtilde_i \leq \bxtilde\}$ converges to a Gaussian process in $\bxtilde$. Using that fact and $\Tschat(\bxtilde) \rightarrow_p \Tsc(\bxtilde)$ it can be shown that 
\begin{align}
\int (\Tschat(\bxtilde) - \Tsc(\bxtilde))\trans d\left\{\nnhalf\sumin(\by_i - \bmuhat_i) I\{\bxtilde_i \leq \bxtilde\}\right\} = o_p(1).
\end{align}
This means that $\bqhat$ and $\bqtilde$ are asymptotically equivalent, and we can examine the limiting behavior of $\bqtilde$ to understand the limiting behavior of $\bqhat$. 

Now, ignoring the intercept $\balpha_0$ for notational simplicity,
\begin{align}
\bqtilde &= \nnhalf\sumin (\by_i - \bmuhat_i)\trans\Tsc_i\\
&= \nnhalf\sumin (\by_i - X_i\balpha)\trans\Tsc_i - \nnhalf\sumin (\balphahat - \balpha)\trans X_i\trans\Tsc_i\\
&= \nnhalf\sumin (\by_i - X_i\balpha)\trans\Tsc_i - \nnhalf\sumin \Usc_i(\balpha)\trans n\inv\sumin(X_i\trans\Tsc_i) + o_p(1)\\
&= \nnhalf\sumin \left\{(\by_i - X_i\balpha)\trans\Tsc_i - \Usc_i(\alpha)\trans E(X_i\trans\Tsc_i)\right\} + o_p(1)\\
&= \nnhalf\sumin\Qsc_i + o_p(1)
\end{align}
where $\Usc_i(\balpha)$ corresponds to the influence function for $\alphahat$ from a null parametric model and $\Qsc_i = (\by_i - X_i\balpha)\trans\Tsc_i - \Usc_i(\alpha)\trans E(X_i\trans\Tsc_i)$. Thus, clearly $\bqtilde$ has a limiting normal distribution by the central limit theorem, which further suggests that the asymptotic distribution of $\Qhat$ is a mixture of $\chi_1^2$s where the mixing probabilities are given by the eigenvalues of $\cov(\Qsc_i)$.

\subsection{The voom procedure}

Let $r_{ijt}$ be the read count from individual $i$ for gene $j$ at time $t$. $L_i=\sum_{j=1}^Gr_{ijt}$ is then the library size, that can vary from one sample to another. \cite{Law2014} propose the following transformation to obtain log-counts per million value from sample $i$ for gene $g$:

\begin{align}
y_{ijt} = \log_2\left(10^6\dfrac{0.5+r_{ijt}}{1+L_i}\right)
\end{align}

Counts are offset away from zero by 0.5 to avoid taking the log of zero (and this also reduces the variability of log-cpm for low expression genes) while the library size is offset by 1 to ensure that any log-cpm is below 1 million. Let $X$ be the design matrix of factors that are suspected to influence gene expression.

The voom procedure to estimate mean-variance weights proposed by \cite{Law2014} is the following:

\begin{enumerate}
 	\item $\widehat{\alpha}$ is the OLS estimate from the linear model: $y_{ijt}=  \bx_i^T\alpha + \varepsilon_{ijt}$ with $\varepsilon_{ijt}\sim\mathcal{N}(0,\sigma_j)$
 	\item Let $\displaystyle \widehat{s}_j=\sqrt{\sum_{i=1}^n\left(y_{ijt}-\bx_i^T\widehat{\alpha}\right)^2}$
 	\item Let $\displaystyle\tilde{r}_j =\frac{1}{n}\sum_{i=1}^n y_{ijt} +\log_2\left(\prod_{i=1}^n(1+\sum_{g=1}^Gr_{ijt})\right)^{1/n}-\log_2(10^6)$ be the average log-count value.
	\item $\widehat{f}(\cdot)$ is the predictor obtained from the LOWESS regression \citep{Cleveland1979} of $\sqrt{\widehat{s}_j}$ over $\tilde{r}_j$. The final precision weights are then defined as follows:
    
\begin{align}
w_{ijt}=\left[\widehat{f}\left(\bx_i^T\widehat{\alpha} + \log_2(1 + \sum_{g=1}^Gr_{ijt}) -\log_2(10^6)\right)\right]^{-4}
\end{align}
\end{enumerate}
\bigskip


\subsection{Additional simulations}

\begin{figure}[!h]
\centering
	\includegraphics[width=0.6\textwidth]{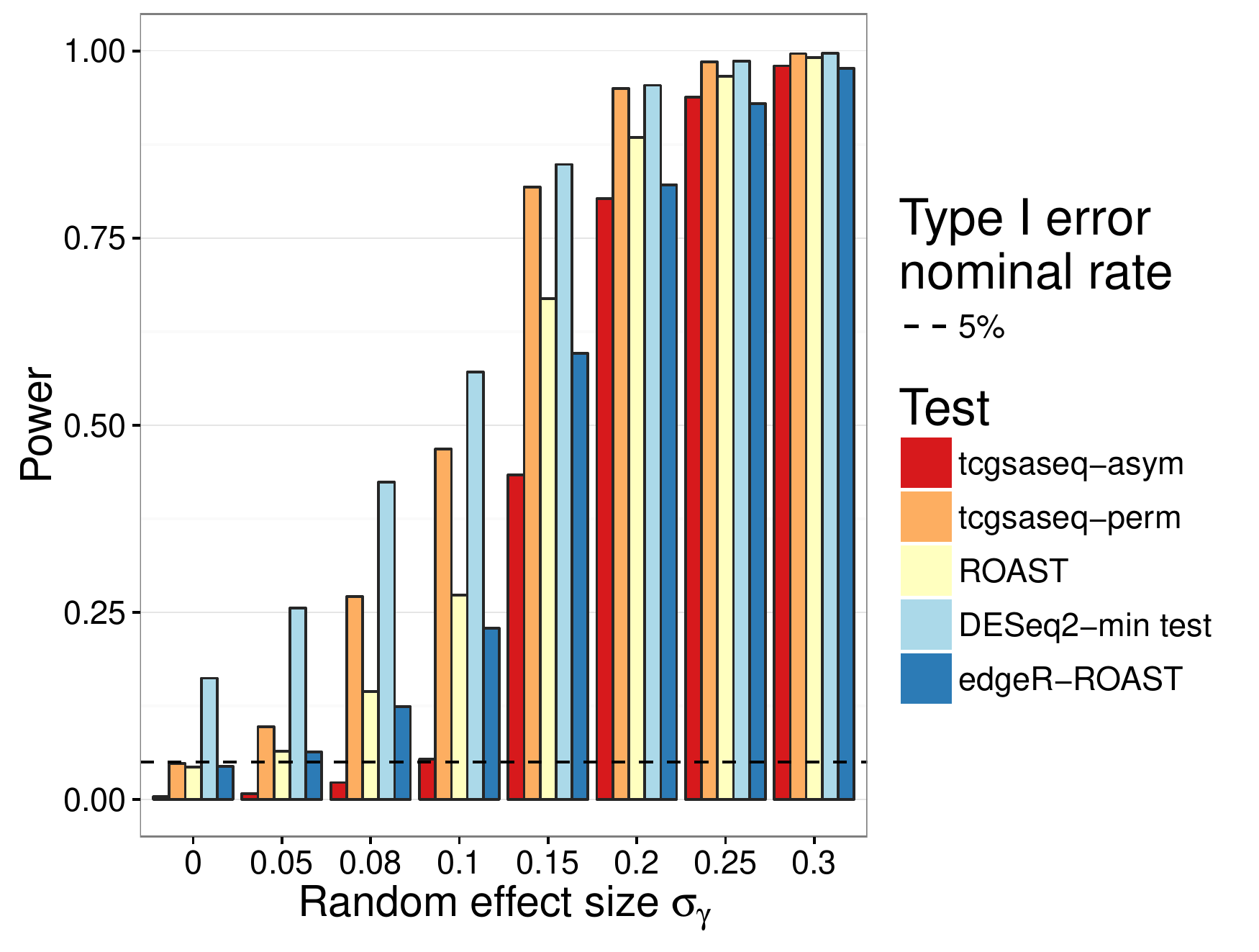}
    \caption{Power evaluation for heterogeneous gene sets in realistically simulated data with a small sample size, based on 500 simulations.}
    \label{power_voomlike_2}
\end{figure}

Here we provide additional simulation results in the realistic setting of a small sample size. We generated data for 6 individuals, each measured at 3 time points according to the scenario described in \cite{Law2014}, using the script provided in their Supplementary Material:
\begin{align}
y_{ijt} = \log\left(\frac{(a_{ij}+0.5) 10^6}{\sum_j a_{ij} +1}\right) + \gamma_{j} t_{it} + \epsilon_{ijt}
\end{align}
where $t_{it} \in \{1, 2, 3\}$, $\epsilon_{ijt} \sim N(0, 0.05)$, $a_{ij}\sim Pois(\lambda_{ij})$, $\lambda_{ij}\sim Gamma(\frac{\kappa_{ij}}{40 m_{ij}},$ $\frac{{40 m_{ij}}}{\kappa_{ij} \mu_{ij}})$, $\kappa_{ij}\sim\chi^2(40)$, $m_{ij}=0.2 +1/\sqrt{\mu_{ij}}$, $\gamma_j\sim N(0,\sigma_\gamma)$, and $\mu_{ij}$ follows an empirical baseline distribution derived from real data, provided in supplementary material of \cite{Law2014}. We set $n=18$ with $n_i=3$, and gene sets were constructed such that $p \in (30, \dots, 400)$ and for every gene pair $(j, j')$ in the set $cor(\ba_{j}, \ba_{j'})>0.8$. $\sigma_\gamma)$ varied among $(0, 0.02, 0.05, 0.1, 0.15, 0.2, 0.25, 0.3)$. The results are very similar to those for homogeneous gene sets, see Figure \ref{power_voomlike_2}.

\bibliographystyle{biorefs}
\bibliography{TcGSAseq}

\end{document}